\documentstyle[editedvolume,psfig]{crckapb}
\newcommand{\msun}{\mbox{${\rm M}_\odot$}}

\begin{opening}
\title{Formation of Low-mass Black Hole X-ray Transients}

\author{Simon Portegies Zwart}
\institute{Astronomical Institute {\em Anton Pannekoek}, Amsterdam,
	   The Netherlands} 
\author{Frank Verbunt}
\institute{Astronomical Institute Utrecht, The Netherlands}
\author{Ene Ergma}
\institute{Tartu University, Tartu, Estonia}
\end{opening}
\runningtitle{The formation of Black Hole X-ray transients}
\begin{document}
\section{Introduction}
We study the formation of low-mass X-ray binaries with
a black hole as accreting object. 
The performed semi-analytic analysis
reveals that the formation rate of black holes in low-mass X-ray
binaries is about two orders of magnitude smaller than that of systems
with a neutron star as accretor. This is contradicted by the
six observed systems, which are all transients, which suggest that the
majority of these systems has not been seen jet.
The birthrate for both type of objects are expected to be similar
(for reviews see Cowley 1992, Tanaka \& Lewin 1995).

Solution of this conundrum 
requires either that stars with masses less than 20~\msun\ can
evolve into black holes, or that stellar wind from a member of
a binary is accompanied by a much larger loss of angular momentum
than hitherto assumed. 

The most popular formation scenario for low-mass X-ray binaries
proposes a relatively wide binary with an extremely small mass ratio as the
progenitor system (van den Heuvel 1983).  The massive star evolves to fill
its Roche lobe, engulfs its low-mass companion followed by a spiral-in.  
A close binary remains if the spiral-in ceases before
the low-mass companion coalesces with the compact helium core of the
primary.  The helium core continues its evolution and may turn into a
neutron star or a black hole. 

\section{The maximum initial separation for Roche-lobe contact}
In the standard scenario the formation of a low-mass X-ray binary
starts with a high-mass primary $M_\circ$ and a low mass secondary
star ($m_\circ \sim 1 \msun$) in a detached binary with semi-major
axis $a_\circ$; the initial mass-ratio $m_\circ/M_\circ$ is
consequently fairly small. 
For simplicity we assume that the initial orbit is circular.  As the
primary evolves and expands on the giant branch it might lose a
considerable fraction of its mass in a stellar wind before the primary
fills its Roche-lobe. This results in an expansion of the orbit
according to the Jeans approximation of conservation of angular
momentum 
(see van den Heuvel 1983): $ a(M + m) = a_\circ(M_\circ + m_\circ)$.
The initial orbital separation $a_\circ$ follows from
the assumption that the primary with mass $M$ ($<M_\circ$) and radius
$R$ ($R\gg R_\circ$) just fits the size of its Roche lobe (Eggleton 1983). 

For each stellar evolution track in the tables of Schaller et al.~(1992) we
can now compute the maximum to the {\em initial} semi-major axis
$a_{\rm max}$ for which the star fills its Roche lobe. Note that mass
loss by the stellar wind might cause 
$a_{\rm max}$ to be reached at an earlier epoch than at the maximum radius of
the primary (see Fig.~2 in Portegies Zwart et al.~1997).

\section{The minimum initial separation to survive mass transfer}
Roche-lobe contact to the small mass companion causes mass transfer
to be highly unstable and a common-envelope phase cannot be avoided.
A close binary remains after the spiral in provided that the primary's
envelope is fully ejected before the secondary star coalesces with the
compact core (with mass $M_{\rm He} = 0.073M_\circ^{1.42}$) of the
primary. 
The minimum orbital separation {\em after} the spiral in is
provided by the secondary with mass $m$ ($\equiv m_\circ$)
and radius $r$ ($\equiv r_\circ$) to fill its Roche-lobe. The
core of the primary is more massive than its companion and its size is
smaller. The secondary therefore gives the strongest limit.

The reduction in the semi-major axis during the spiral-in can be
computed by comparing the binding energy of the primary's envelope
with the orbital binding-energy of the binary (Webbink 1984): 
$ \Delta E_{\rm env} = \alpha_{ce} \Delta E_{\rm orb}$.  
Here $\alpha_{ce}$ gives the
efficiency in which the energy generated by the friction between the
secondary and the common envelope will blow off the envelope at the
cost of orbital energy.  The minimum semi-major axis $a_{\rm min}$ for
which the binary survives spiral-in is computed accordingly.

\begin{figure}
\centerline{
\psfig{file=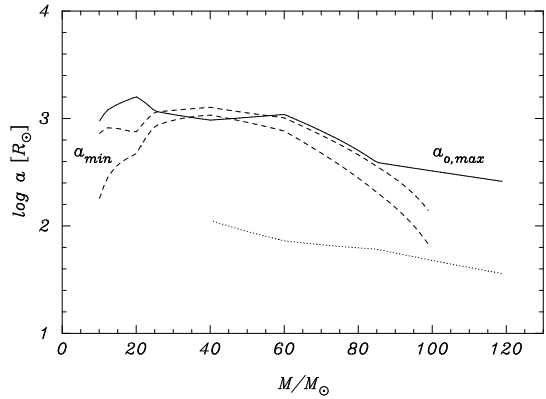,bbllx=570pt,bblly=40pt,bburx=110pt,bbury=690pt,height=5cm,angle=-90}}
\caption{Lower limit to the initial semi-major axis at which the
binary survives the spiral-in, as function of the initial mass of the
primary, calculated with use of the evolutionary sequences by
Schaller et al.\ (1992).
The upper (lower) dashed line gives the limit determined from the
condition that the secondary star (helium core) is smaller than its 
Roche-lobe, 
The solid line gives the upper limit at which the primary 
reaches its Roche lobe at its maximum radius.
The secondary is assumed to be a 1~\msun star.
The dotted line indicates the initial semi-major 
axis of a binary in which a 1~\msun\ secondary can fill its Roche lobe
after the primary has lost its
entire envelope without ever having filled its Roche lobe
}
\label{survival}\end{figure}

From the orbital separation that corresponds barely to survival of the
common-envelope we can compute the {\em initial} orbital separation by
applying Jeans approximation for the phase of mass loss before
Roche-lobe contact.
The masses and radii of the primary, required for this study, are 
taken from stellar evolution tracks of Schaller et al.~(1992).
Figure~\ref{survival} gives the minimum and maximum to the initial
semi-major axes.  
There where the solid line is above the dashed line primaries are able to
fill their Roche-lobe and survive the spiral-in.
The small surface area enclosed by the solid line
($a_{\circ, \rm max}$) and the upper dashed line ($a_{\rm min}$) gives the
allotted region for the formation of a low-mass X-ray binary.

\subsection{Relative birthrate}
The formation rate of binaries that reach
Roche-lobe contact and survive the spiral-in
can be estimated by computing the birth rate of binaries with 
suitable initial parameters; primary mass, mass ratio and semi-major
axis.
For simplicity we neglect the eccentricity.
For the initial mass function for the primary 
we use a Salpeter function (Salpeter 1964).
The initial semi-major axis distribution 
is taken flat in $\log a$ and the initial mass-ratio
distribution is taken uniform between 0 and 1.

\begin{figure}
\centerline{
\psfig{file=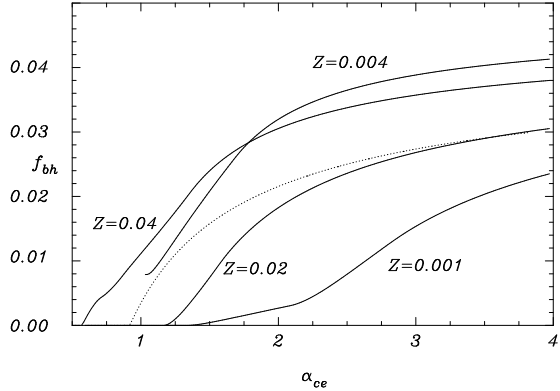,bbllx=570pt,bblly=40pt,bburx=110pt,bbury=690pt,height=5cm,angle=-90}}
\caption{The binaries with a black hole as a
fraction of the neutron--star binaries (X-axis) as a function of the
common-envelope parameter $\alpha_{ce}$. The different lines give the
result for stellar evolution models with different metalicities:
$Z=0.001$ (Schaller et al.\ 1992), $Z=0.004$ (Charbonnel et al.\
1993), $Z=0.02$ (solid: Maeder \& Meynet 1988, dotted: Schaller et
al.\ 1992) and $Z=0.04$ (Schaerer et al.\ 1993).  
}
\label{figfbh_ns}
\end{figure}

Primaries with an initial mass between 10~\msun\ and
40~\msun\ are assumed to leave a neutron star after
the supernova,  
progenitors with a mass between 40~\msun\ and 
100~\msun\ form black holes.
The secondary star is assumed to have a mass between $0.85$~\msun\ and
1.15~\msun. 
The minima and maxima for the initial semi-major axis are
computed as described in the previous section.
We integrate the initial distribution functions for the primary mass,
the mass ratio and the semi-major axis between the appropriate mass 
limits and the corresponding limits for the semi-major axis 
$a_{\rm min}$ and $a_{\circ, \rm max}$ (see Fig.~\ref{survival}):
With the proper normalization for the initial mass function 
and a binary fraction of 50\%
the galactic formation rate of binaries that reach Roche-lobe
contact and survive the spiral-in is
$2.2\cdot10^{-6}\, {\rm yr}^{-1}$ for the binaries with a neutron
star (which is comparable to the estimates from observations)
and $9.6\cdot10^{-9}\, {\rm yr}^{-1}$ for the
black-hole binaries (see Portegies Zwart et al.~1997). 
The formation rate of low-mass X-ray binaries with a black hole is
about 1\%\ of the formation rate of low-mass X-ray binaries with a
neutron star, whereas the observations indicate equal formation rates
for these two types of binaries.
Figure~\ref{figfbh_ns} gives the relative birthrates of low-mass X-ray
binaries with a black hole relative to those with a neutron star for
various metalicities ranging from $Z=0.001$ to $Z=0.04$ as a function
of the common envelope efficiency parameter $\alpha_{ce}$.

\section{Conclusions}
Our computations predict a formation rate for low mass X-ray binaries
with a black hole which is
much smaller than the value derived from the observed numbers and
estimated X-ray lifetime.
The discrepancy between theoretical and observed formation rates
cannot be solved by invoking different metalicities for the
progenitor systems, nor by assuming different efficiencies
for the envelope rejection during spiral-in.

Black-hole binaries can be produced in larger numbers only if
we assume that stars with initial masses less than 
$\sim 20$~\msun\ collapse in to black holes; or alternatively if 
it is assumed that the angular momentum loss caused by the
stellar wind is so high that the binary orbit shrinks.

\acknowledgements
This work was supported in part by the Netherlands
Organization for Scientific Research (NWO) under grant PGS 13-4109 and
by the Leids Kerkhoven Boscha Fonds. 
This investigation is supported by NWO under Pioneer grand PGS~78-277
to F.~Verbunt and by Spinoza grant 08-0 to E.~P.~J.~van den Heuvel.


\begin{thebibliography}{}
\bibitem[]{} Charbonnel, C., Meynet, G., Maeder, A.~S., Schaller, G.,
	Schaerer, D. 1993, A\&AS, 101, 415 
\bibitem[]{}Cowley, A. 1992, ARA\&A, 30, 287
\bibitem[]{} Maeder, A., Meynet, G. 1989, A\&A, 210, 155
\bibitem[]{} Salpeter, E.~E. 1964, ApJ, 140, 796
\bibitem[]{} Schaerer, D., Charbonnel, C., Meynet, G., Maeder, A.~S.,
	Schaller, G. 1993,  A\&AS, 102, 339
\bibitem[]{} Schaller, G., Schaerer, D., Meynet, G., Maeder, A.~S. 1992,
	A\&AS, 96, 269
\bibitem[]{} Portegies Zwart, S.\ F., Verbunt, F., Ergma, E., 1997, A\&A 
	321, 207
\bibitem[]{} Tanaka, Y., Lewin, W. H.~G. 1995, in W.~H. G. e.~a. Lewin
	(ed.), X-ray binaries, Cambridge University  Press,  536
\bibitem[]{}van~den Heuvel, E. 1983, in W. Lewin, E. van~den Heuvel
	(eds.), Accretion-driven stellar X-ray sources, Cambridge
	U.P., Cambridge, p.~303 
\bibitem[]{}van~den Heuvel, E.~P.~J., Habets, G. 1984, Nat, 309, 598
\bibitem[]{}Webbink, R.~F. 1984, ApJ, 277, 355
\end{thebibliography}
\end{document}